\documentclass[11pt]{article}
\usepackage[dvipsnames,table]{xcolor}
\usepackage[a4paper, left=1in, right=1in, top=1.2in, bottom=1in]{geometry}
\usepackage{setspace}
\usepackage{authblk}
\usepackage[font=footnotesize]{caption}
\usepackage[style=ieee, citestyle=numeric-comp, backend=biber, maxbibnames=1000]{biblatex}

\usepackage{xurl}

\usepackage[T1]{fontenc}
\usepackage[final]{microtype}
\usepackage{listings}
\usepackage{tikz}
\usetikzlibrary{shapes, arrows, shadows, backgrounds, fit, positioning}
\pgfdeclarelayer{background}
\pgfdeclarelayer{middle}
\pgfsetlayers{background,middle,main}
\usepackage{quantikz}
\usepackage{todonotes}
\usepackage{booktabs}
\usepackage{tabularx}
\usepackage{amsmath}
\usepackage[T1]{fontenc}

\definecolor{lightgray}{rgb}{0.9,0.9,0.9}
\definecolor{mediumgray}{rgb}{0.8,0.8,0.8}

\DeclareCaptionType{listing}[Listing][List of Listings]

\everymath=\expandafter{\the\everymath\displaystyle}

\lstdefinestyle{mystyle}{
	backgroundcolor=\color{white},
	commentstyle=\color{Green},
	keywordstyle=\color{blue},
	stringstyle=\color{magenta},
	numberstyle=\scriptsize\color{gray},
	basicstyle=\footnotesize,
	xleftmargin=2em,
	breaklines = true,
	captionpos = b,
	keepspaces = true,
	numbers = left,
	numbersep = 7pt,
	showspaces = false,
	showstringspaces = false,
	showtabs = false,
	tabsize = 2,
	frame=single,
	framexleftmargin=1.5em,
	xleftmargin=2em,
}
\begin{document}
\title{A Survey on Integrating Quantum Computers into High Performance Computing Systems}

\author[]{Philip Döbler}
\author[]{Manpreet Singh Jattana}

\affil[]{Modular Supercomputing and Quantum Computing, Goethe University Frankfurt, Kettenhofweg 139, 60325 Frankfurt am Main, Germany\\
\texttt{\{doebler, jattana\}@em.uni-frankfurt.de}}

\date{July 4, 2025}

\maketitle

\begin{abstract}
    \noindent
    Quantum computers use quantum mechanical phenomena to perform conventionally intractable calculations for specific problems. Despite being universal machines, quantum computers are not expected to replace classical computers, but rather, to complement them and form hybrid systems. This makes integrating quantum computers into high performance computing (HPC) systems an increasingly relevant topic. We present a structured literature review on the integration aspect. We methodologically search literature databases and manually evaluate 107 publications. These publications are divided into seven categories that describe the state of the art in each category. After a brief quantitative analysis of the literature, this survey deals with the hardware architecture of hybrid quantum-classical systems, as well as the software stack. We observe the development of a wide range of tools enabling hybrid systems and emphasize the need for future standardization of interfaces and methods to foster synergy.
\end{abstract}

\section{Introduction}
\label{introduction}

In recent decades, high performance computing (HPC) has led to countless breakthroughs in science and technology. This was made possible by the continuous increase in computing power, which has been driven primarily by the progressive miniaturization of semiconductors, resulting in an increasing number of transistors on a single chip. This phenomenon is known as Moore's Law~\cite{moores_law}. When further increases in clock rates became difficult, multicore processors and systems were developed to increase computing power. This was followed by the development of specialized accelerators such as graphics processing units (GPUs). However, achieving further speedup is becoming increasingly difficult. At some point, Moore's Law is projected to come to an end when structure sizes reach atomic limits.

One way to further increase computing power is to use quantum computers. They can solve some problems that are intractable using classical methods by leveraging quantum mechanical phenomena~\cite{shor,grover}. However, many algorithms cannot be accelerated by quantum computers. Therefore, it makes sense to combine classical HPC systems and quantum computers. Another reason to host quantum computers in HPC centers is their extensive experience in providing computing resources to users. Currently, quantum computers are expensive and complex to operate. Therefore, it is sensible to use the available resources as efficiently as possible and give as many users as possible access to them. This integration benefits both the quantum computing and HPC communities. It enables hybrid algorithms and allows quantum computing specialists to focus on their research without worrying about setting up and maintaining the infrastructure. For the HPC community, quantum computers offer the possibility of further increasing their systems' computing power and expanding the pool of potential users. Experience gained from integrating quantum computers could be relevant for integrating other exotic technologies, such as neuromorphic or analog computers.

To provide an overview of the growing number of publications on integrating quantum computers into HPC systems, this paper presents a structured literature review on the topic. We search literature databases using defined keywords and manually sort the results into different categories. We then discuss the individual categories on the basis of the literature found. We have three goals: First, readers should be able to gain a quick and comprehensive overview. Second, we aim to present the state of the art in order to identify gaps and starting points for further research. Third, we aim to provide a practical use for people who are working on the integration of quantum computers into HPC systems.

The rest of the paper is structured as follows: Section~\ref{sec:scope_approach} describes the scope of our literature review and our approach. Section~\ref{sec:quantitative_analysis} contains a quantitative analysis of the literature about integration of quantum computers into HPC systems. The subsequent sections present the results of the literature review. First, we present publications that provide a general overview of the topic (Section~\ref{sec:overview}). Then, we discuss the hardware architecture (Section~\ref{sec:hardware}) and software stack (Section~\ref{sec:software}), followed by a section on benchmarks (Section~\ref{sec:benchmarking}). We provide a discussion about the state of the art and future topics in Section~\ref{sec:future_topics} and conclude the paper in Section~\ref{sec:conclusion}.

\section{Scope and Approach}
\label{sec:scope_approach}

The scope of this work are scientific publications that focus on the integration of quantum computers into HPC systems. We do not consider literature that focuses on only one of the aspects. We also do not consider literature that deals with quantum computing and HPC without a focus on the integration of quantum computers into HPC systems. For example, there is extensive literature on the simulation of quantum computers on HPC systems. This is not our focus, as in this case HPC is a tool for quantum computing research, but the literature does not address the actual interplay between the two.

To find literature, we use the literature database Web of Science~\cite{wos} and the preprint repository arXiv~\cite{arxiv}. Web of Science contains literature from various journals and conference proceedings. arXiv contains preprints of some articles published in peer-reviewed journals or at conferences, as well as e-prints not published elsewhere. Some literature is not available as preprints, while others are currently only published on arXiv. We therefore use both databases to find as many publications as possible. Naturally, literature in the desired area often contains the words ``HPC'' and ``quantum computer'' or synonyms in their titles, keywords or abstracts. In addition, there are papers that contain the word ``heterogeneous'' in combination with ``quantum computer'', ``quantum hybrid'' or similar words. For the structured search, we therefore defined the keyword groups in Table~\ref{tab:literature_research_keywords}. The results are found in the following way:
\begin{equation}
    \text{Results} = (\text{Group 1a} \land \text{Group 2a}) \lor (\text{Group 1b} \land \text{Group 2b}).
\end{equation}

\begin{table}[htbp]
    \caption{Keywords for the literature research.}
    \centering
    \begin{tabular}{|l|l|}
        \hline
        Group 1a & Group 2a \\
        \hline
        hpc & qpu\\
        high performance computing & quantum processor\\
        high performance computer & quantum computing\\
        supercomputing & quantum processing unit\\
        supercomputer & quantum computer\\
        \hline
        \hline
        Group 1b & Group 2b\\
        \hline
        heterogeneous & quantum classical\\
        & quantum hybrid\\
        & quantum accelerator\\
        & quantum co processor\\
        & quantum computer\\
        \hline
    \end{tabular}
    \label{tab:literature_research_keywords}
\end{table}

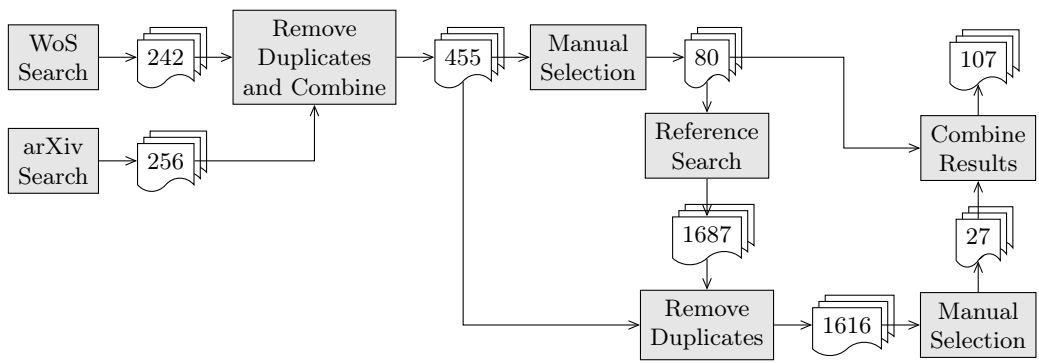
\begin{figure}[htp]
    \centering
    \begin{tikzpicture}[node distance=0.5, every node/.style={font=\footnotesize}]
\node (wos_search) at (0, 0) [rectangle, fill=lightgray, align=center, draw=black] {WoS\\Search};
\node (arx_search) [rectangle, fill=lightgray, align=center, draw=black, below=of wos_search] {arXiv\\Search};
\node (wos_results) [double copy shadow, tape, draw, tape bend top=none, draw=black, fill=white, right=of wos_search] {242};
\node (arx_results) [double copy shadow, tape, draw, tape bend top=none, draw=black, fill=white, right=of arx_search] {256};
\node (remove_duplicates_1) [rectangle, fill=lightgray, align=center, draw=black, right=of wos_results] {Remove\\Duplicates\\and Combine};
\node (comb_results_1) [double copy shadow, tape, draw, tape bend top=none, draw=black, fill=white, right=of remove_duplicates_1] {455};
\node (man_class_1) [rectangle, fill=lightgray, align=center, draw=black, right=of comb_results_1] {Manual\\Selection};
\node (man_res_1) [double copy shadow, tape, draw, tape bend top=none, draw=black, fill=white, right=of man_class_1] {80};
\node (ref_search) [rectangle, fill=lightgray, align=center, draw=black, below=of man_res_1] {Reference\\Search};
\node (ref_search_results) [double copy shadow, tape, draw, tape bend top=none, draw=black, fill=white, below=of ref_search] {1687};
\node (remove_duplicates_2) [rectangle, fill=lightgray, align=center, draw=black, below=of ref_search_results] {Remove\\Duplicates};
\node (comb_results_2) [double copy shadow, tape, draw, tape bend top=none, draw=black, fill=white, right=of remove_duplicates_2] {1616};
\node (man_class_2) [rectangle, fill=lightgray, align=center, draw=black, right=of comb_results_2] {Manual\\Selection};
\node (man_res_2) [double copy shadow, tape, draw, tape bend top=none, draw=black, fill=white, above=of man_class_2] {27};
\node (combine) [rectangle, fill=lightgray, align=center, draw=black, above=of man_res_2] {Combine\\Results};
\node (final_res) [double copy shadow, tape, draw, tape bend top=none, draw=black, fill=white, above=of combine] {107};

\draw[-angle 45] (wos_search) -- (wos_results);
\draw[-angle 45] (arx_search) -- (arx_results);
\draw[-angle 45] (wos_results) -- (remove_duplicates_1);
\draw[-angle 45] (arx_results) -| (remove_duplicates_1);
\draw[-angle 45] (remove_duplicates_1) -- (comb_results_1);
\draw[-angle 45] (comb_results_1) -- (man_class_1);
\draw[-angle 45] (man_class_1) -- (man_res_1);
\draw[-angle 45] (man_res_1) -- (ref_search);
\draw[-angle 45] (ref_search) -- (ref_search_results);
\draw[-angle 45] (ref_search_results) -- (remove_duplicates_2);
\draw[-angle 45] (remove_duplicates_2) -- (comb_results_2);
\draw[-angle 45] (comb_results_2) -- (man_class_2);
\draw[-angle 45] (man_class_2) -- (man_res_2);
\draw[-angle 45] (man_res_2) -- (combine);
\draw[-angle 45] (combine) -- (final_res);
\draw[-angle 45] (comb_results_1) |-  (remove_duplicates_2);
\draw[-angle 45] (man_res_1) -| +(1.8,-0.8) |- (combine);

\end{tikzpicture}
    \caption{Overview of our literature research approach. We query the Web of Science (WoS) database and arXiv for initial results classified by title and abstract. In addition, we check the references of these publications to find further relevant literature. Since most of these references either deal with different topics or are already on our list, we only use 27 out of 1,616.}
    \label{fig:search_approach}
\end{figure}

The process of the literature search is shown in Figure~\ref{fig:search_approach}. We query Web of Science and arXiv with our defined keywords. This search was performed on March~11, 2025 and therefore does not include literature published after that date. In a first step, we remove duplicates and combine the results. It is interesting to note that there is not much overlap between the two databases. We decide whether the papers belong to the desired research area, based on their titles and abstracts. For the resulting 80 papers, we search for all literature cited by them. The results are filtered, which gives us another 27 papers for a total of 107 results.

Based on the full text of the found publications, we identify the following relevant topics that are reflected in the structure of the following document. We take into account that a single publication may cover more than one of these topics.
\begin{itemize}
    \item Overview: These publications provide a general introduction to the topic. This includes early work that outlines general concepts and describes existing systems. We also list related work to our survey here.
    \item Hardware architecture: Publications about the hardware setup of hybrid quantum-classical systems.
    \item Software stack: Publications that describe the software stack that is required for hybrid systems, with the following sub-topics:
    \begin{itemize}
        \item Applications
        \item Programming models and frameworks
        \item Middleware
    \end{itemize}
    \item Benchmarking: Publications about benchmarks designed for hybrid systems.
\end{itemize}
\section{Quantitative Analysis}
\label{sec:quantitative_analysis}

\begin{figure}[htp]
    \centering
    \input{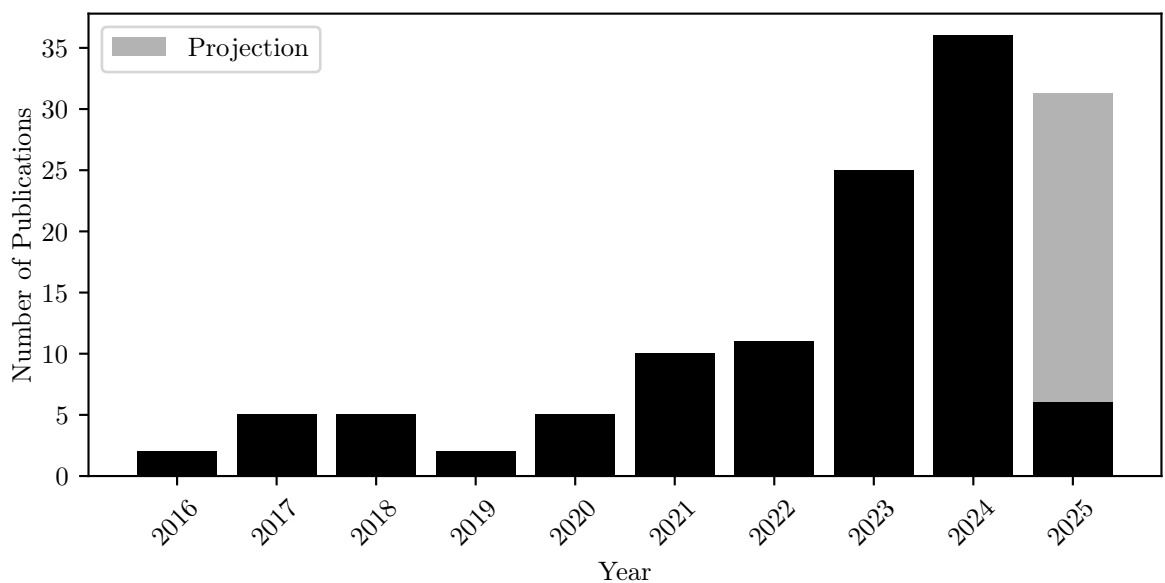}
    \caption{Number of publications per year. For 2025, we estimate the number of publications based on publications in 2025 to date.}
    \label{fig:overview_years}
\end{figure}

We use the bibliometric data to perform some quantitative analysis of the research field. Figure~\ref{fig:overview_years} shows the number of publications per year. We see that the field has gained a lot of traction since the beginning of 2023. For 2025, we project the number of publications by dividing the number of actual publications by the number of days that have already passed and multiplying by the number of days per year.

\begin{figure}[htp]
    \centering
    \input{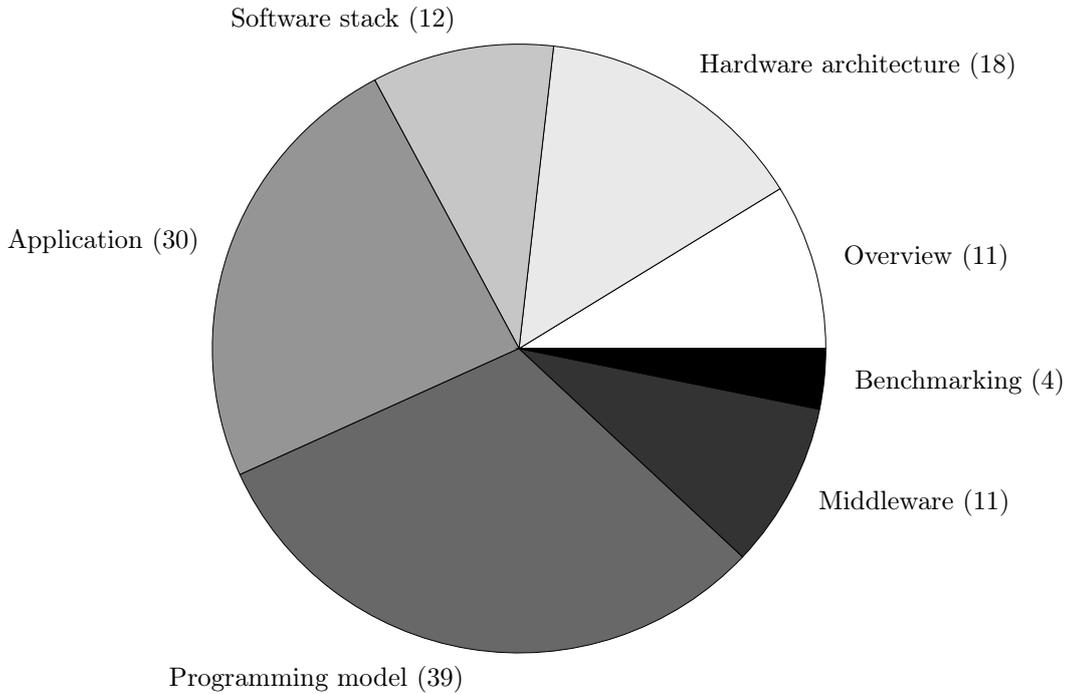}
    \caption{Overview of the different categories and how many publications are available for each.}
    \label{fig:overview_categories}
\end{figure}

Figure~\ref{fig:overview_categories} gives an overview of how many papers are published on each topic. Note that a paper can have multiple topics. Most papers are about programming models, applications, and hardware architecture. Notably, 10 out of 11 publications on middleware (see Section~\ref{sec:middleware}) were only published in 2023 and 2024, despite the central role of middleware in integrating quantum computers into HPC systems. The other topics were published in different years, with no particular accumulation. Work on hardware, software, programming models, and applications was carried out in parallel.

\section{Overview}
\label{sec:overview}

Several publications provide an overview of the subject. One of the earliest publications on the subject, from 2017, predicts the impact quantum computers will have on HPC~\cite{literature_res_196}. The authors argue that quantum computers will serve as accelerators rather than standalone machines and advocate for the development of standardized software stacks and hardware abstraction layers. Reference~\cite{literature_res_33} performs a bibliometric analysis of HPC and quantum computing. Unlike us, the authors consider all publications about HPC or quantum computing. The 3,435 papers they found are analyzed based on their bibliometric data, but not manually based on the content. In contrast, Reference~\cite{literature_res_16} is limited to publications that address the integration of quantum computers into HPC systems. Basic concepts of hardware and software integration are presented, but no structured literature review is conducted. A path to scaling from hundreds to millions of qubits, along with the associated challenges, is presented in Reference~\cite{literature_res_274}. One section of the paper is devoted to hybrid quantum-classical computing. It outlines solutions for a hybrid software stack, hybrid applications, and benchmarking of hybrid systems.

The goal of HPC center operators is to ensure system dependability. For systems that integrate quantum computers, there are particular security, privacy, reproducibility, and resilience challenges, as discussed in Reference~\cite{literature_res_302}. For example, reproducibility is more difficult to guarantee due to quantum noise, and securing the system becomes more difficult as it becomes more heterogeneous. Reference~\cite{literature_res_376} also focuses on reliability. With a photonic quantum computer integrated into an HPC system, an availability of 92\% was achieved over six months. While this is low compared to classical HPC hardware, it should be noted that there was no redundancy in this system and that it required regular recalibration, which was counted as downtime.

For a scientific topic to be successfully applied, it must be accessible to the general public. Traditionally, quantum computing lectures stem from quantum physics and are therefore more physics-oriented than computer science-oriented. Reference~\cite{literature_res_15} addresses this gap in computer science curricula by presenting a series of lectures and exercises specifically for hybrid quantum-classical systems. Since this lecture does not assume knowledge of quantum physics, it begins with the fundamentals of quantum computation. Then, it focuses on variational algorithms and quantum machine learning as examples of hybrid algorithms. This aligns well with the types of hybrid applications identified in Section~\ref{sec:applications}. The course concludes with lectures on photonic quantum computing and quantum error mitigation.

Several publications describe in general the hardware and software layers necessary for quantum accelerators~\cite{literature_res_206,literature_res_179,literature_res_457}. They all have one thing in common: programs are compiled into a quantum instruction set architecture (QISA). Hybrid programs are divided into classical and quantum parts. A special compiler handles the quantum part and takes care of quantum-specific tasks, such as routing and transpilation into the native gate set. During runtime, the instructions are loaded from main memory, decoded, and executed on the quantum computer. In the following sections, we take a deeper look into the hardware architecture and software stack for hybrid quantum-classical systems.

\section{Hardware Architecture}
\label{sec:hardware}

This section introduces different concepts for the hardware architecture. We distinguish between macro and micro architecture. Macro architecture describes where the components are located and how they interact at a high level. For example, the quantum computer is shown as a single block without considering the internal components. The micro architecture takes a more detailed look at certain aspects of the system.

\subsection{Definitions}
\begin{figure}[htp]
    \centering
    \begin{tikzpicture}[node distance=0.5, every node/.style={font=\footnotesize}]
\node(qubits) [rectangle, draw] {Qubits};
\node(mce) [rectangle, above=3mm of qubits, draw] {MCE}; 
\node(qcp) [rectangle, above=3mm of mce, draw] {QCP}; 
\node (qpu) [draw, densely dashed, fit={(qubits) (mce)}, inner sep=1mm] {};
\node (qputext) [below=0 of qpu] {QPU};
\node (qcu) [draw, densely dotted, fit={(qcp) (mce)}, inner sep=1mm] {};
\node (qcutext) [above=0 of qcu] {QCU};
\node (qc) [draw, fit={(qpu) (qcu) (qputext) (qcutext)}, inner sep=3mm] {};
\node (qctext) [above=0 of qc] {QC};
\draw (qcp) -- (mce);
\draw (mce) -- (qubits);
\node(other) [right=2cm of qcp] {\ldots};
\node(otherq) [right=1.75cm of qpu] {\ldots};
\draw (qcp) -- (other);
\draw[densely dashed] (qpu) -- (otherq);
\end{tikzpicture}
    \caption{Components of a Quantum Computer (QC). The MCE controls the qubits. Together, they form the QPU. The QCP is a classical processor that handles classical communication. There can also be a quantum interface (dashed line).}
    \label{fig:definitions}
\end{figure}
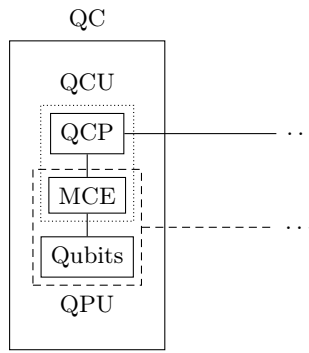

Terms for quantum hardware are used differently in different works. This does not mean that the terms are used incorrectly in these publications. We define and use these terms consistently here, as described below, even if the usage differs in some cases from the work to which we refer. Figure~\ref{fig:definitions} shows the schematic structure of quantum hardware with the following components:
\begin{itemize}
    \item Quantum computer: A quantum computer is a system that can perform computations based on the principles of quantum mechanics. In addition to the qubits themselves, a quantum computer includes all the hardware needed to receive and process jobs through a classical interface, as well as components for cooling, environmental shielding, enclosure, etc., depending on the type of quantum computer.
    \item Quantum control processor (QCP): The QCP is a classical processor that can communicate with other classical processors through a classical interface and is responsible for issuing instructions to the quantum hardware.
    \item Measurement and control equipment (MCE): The MCE is all the hardware needed to manipulate and measure qubits. Depending on the technology, this can for example be an arbitrary waveform generator or a laser.
    \item Qubits: The qubits are a physical realization of a two-level quantum mechanical system, whose quantum states can be prepared, manipulated and measured to perform computations.
    \item Quantum control unit (QCU): The QCU consists of the QCP and MCE.
    \item Quantum processing unit (QPU): The QPU consists of the MCE and the qubits.
    \item Interconnect: The quantum computer is connected with a classical interface via the QCP. An optional quantum interface can be connected to the QPU to allow entanglement across different quantum computers.
\end{itemize}

\subsection{Macro-Architecture}
There is a general consensus in the literature that quantum computers work best as accelerators, because they are good at certain tasks and bad at others. However, there are different concepts for the architecture of this accelerator. Reference~\cite{literature_res_199} introduces the concept of loose versus tight integration. The authors define loose integration as a quantum computer that is a standalone device connected to a classical computer via a network. Tight integration means that the quantum computer is moved closer to the classical hardware and integrated into a common system. The loose integration model can be compared to a client-server model, while the tight integration model is similar to how GPUs are integrated today. 

Reference~\cite{literature_res_458} makes a similar distinction. The authors differentiate between location, which is the relative position of quantum computers and classical hardware, and density, which is the ratio of classical processors to quantum computers shared by those processors. For locality, they distinguish between remote, co-located, and on-node integration. The first two are subtypes of the loose integration model, while on-node integration is identical to the tight integration model.

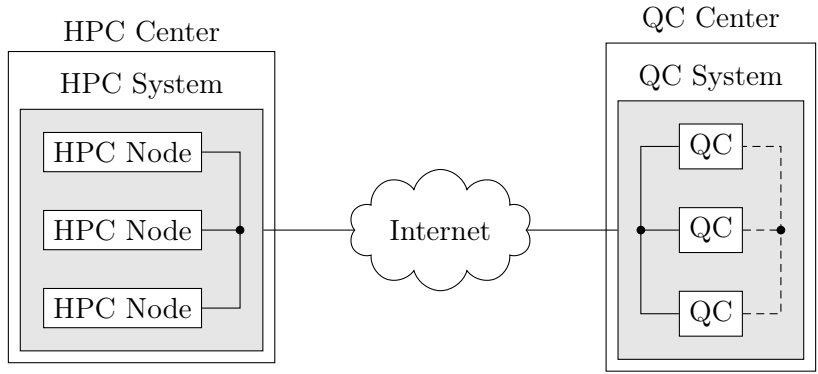
\begin{figure}[htp]
    \centering
    \begin{tikzpicture}
\node(n1) [draw, rectangle, fill=white] {HPC Node};
\node(n2) [draw, rectangle, fill=white, below=5mm of n1] {HPC Node};
\node(n3) [draw, rectangle, fill=white, below=5mm of n2] {HPC Node};
\coordinate(n2r) at ($(n2.east) + (0.5,0)$);
\draw (n3) -| (n2r) |- (n1);
\draw (n2) -- (n2r);
\filldraw[black] (n2r) circle (0.5mm);
\begin{pgfonlayer}{background}
    \node (hpc) [draw, fill=lightgray, fit={(n1) (n3) (n2r)}, inner sep=3mm] {};
\end{pgfonlayer}
\node(hpct)[above=0mm of hpc.north] {HPC System};

\node (internet) [draw, cloud, aspect=2, right=2cm of n2] {Internet};

\node(q2) [draw, rectangle, fill=white, right=2cm of internet] {QC};
\node(q1) [draw, rectangle, fill=white, above=5mm of q2] {QC};
\node(q3) [draw, rectangle, fill=white, below=5mm of q2] {QC};
\coordinate(q2l) at ($(q2.west) - (0.5,0)$);
\coordinate(q2r) at ($(q2.east) + (0.5,0)$);
\filldraw[black] (q2r) circle (0.5mm);
\filldraw[black] (q2l) circle (0.5mm);
\draw [densely dashed] (q3) -| (q2r) |- (q1);
\draw [densely dashed] (q2) -- (q2r);
\draw (q3) -| (q2l) |- (q1);
\draw (q2) -- (q2l);
\begin{pgfonlayer}{background}
    \node (qc) [draw, fill=lightgray, fit={(q1) (q3) (q2r) (q2l)}, inner sep=3mm] {};
\end{pgfonlayer}
\node(qct)[above=0mm of qc.north] {QC System};

\node(hpcc) [draw, fit={(qc) (qct)}, inner sep=1.5mm] {};
\node(qcc) [draw, fit={(hpc) (hpct)}, inner sep=1.5mm] {};

\node[above=0mm of hpcc.north] {QC Center};
\node[above=0mm of qcc.north] {HPC Center};

\draw (hpc) -- (internet);
\draw (qc) -- (internet);
\end{tikzpicture}
    \caption{Schematic of the remote integration model. Classical and quantum hardware are hosted in different locations and connected via the Internet. The quantum computers can optionally be connected with a quantum interconnect (dashed lines).}
    \label{fig:macro_architecture_loose_internet}
\end{figure}

Remote integration means that the quantum computer and classical hardware are hosted in different locations and connected via the Internet (see Figure~\ref{fig:macro_architecture_loose_internet}). This integration model is the lowest barrier for traditional HPC centers to support quantum computing. There is no need to purchase dedicated machines, instead computing time can be purchased from existing hosts. In this way, different technologies can be experimented with at low cost. However, since a public network is used to exchange potentially confidential data, data security must be taken into account. In addition, quantum computers are usually shared by multiple users, which does not allow for efficient scheduling. An example of the remote integration model is the Nordic-Estonian Quantum Computer e-Infrastructure Quest (NordIQuEst) platform~\cite{literature_res_321}, which couples the HPC system LUMI~\cite{lumi} and other classical resources with a five-qubit superconducting quantum computer. Both are hosted in different locations. The software connection between these two systems is described in Section~\ref{sec:middleware}.

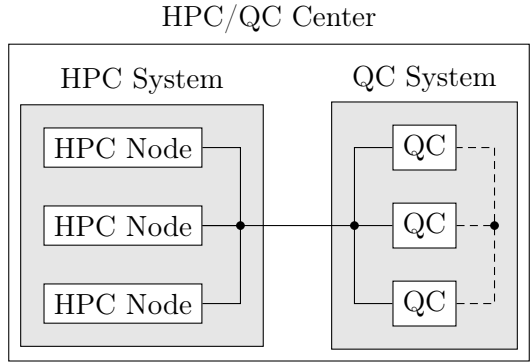
\begin{figure}[htp]
    \centering
    \begin{tikzpicture}
\node(n1) [draw, rectangle, fill=white] {HPC Node};
\node(n2) [draw, rectangle, fill=white, below=5mm of n1] {HPC Node};
\node(n3) [draw, rectangle, fill=white, below=5mm of n2] {HPC Node};
\coordinate(n2r) at ($(n2.east) + (0.5,0)$);
\filldraw[black] (n2r) circle (0.5mm);
\draw (n3) -| (n2r) |- (n1);
\draw (n2) -- (n2r);
\begin{pgfonlayer}{background}
    \node (hpc) [draw, fill=lightgray, fit={(n1) (n3) (n2r)}, inner sep=3mm] {};
\end{pgfonlayer}
\node(hpct)[above=0mm of hpc.north] {HPC System};

\node(q2) [draw, rectangle, fill=white, right=2.5cm of n2] {QC};
\node(q1) [draw, rectangle, fill=white, right=2.5cm of n1] {QC};
\node(q3) [draw, rectangle, fill=white, right=2.5cm of n3] {QC};
\coordinate(q2r) at ($(q2.east) + (0.5,0)$);
\coordinate(q2l) at ($(q2.west) - (0.5,0)$);
\filldraw[black] (q2r) circle (0.5mm);
\filldraw[black] (q2l) circle (0.5mm);
\draw [densely dashed] (q3) -| (q2r) |- (q1);
\draw [densely dashed] (q2) -- (q2r);
\draw (q3) -| (q2l) |- (q1);
\draw (n2) -- (q2);
\begin{pgfonlayer}{background}
    \node (qc) [draw, fill=lightgray, fit={(q1) (q3) (q2r) (q2l)}, inner sep=3mm] {};
\end{pgfonlayer}
\node(qct)[above=0mm of qc.north] {QC System};
\node(center) [draw, fit={(qc) (hpc) (qct) (hpct)}, inner sep=1.5mm] {};
\node(t) [above=0 of center] {HPC/QC Center};

\end{tikzpicture}
    \caption{Schematic of the co-located integration model. The HPC and QC systems are hosted in the same center and connected by a high-bandwidth, low-latency interconnect. The quantum computers can optionally be connected via a quantum interconnect (dashed lines).}
    \label{fig:macro_architecture_loose_on-premise}
\end{figure}

Co-located, also called on-premises, means that the quantum computer and classical hardware are hosted at the same site and connected over a local network, with much lower latency. This approach is illustrated in Figure~\ref{fig:macro_architecture_loose_on-premise}. In addition to low latency, this allows for more sophisticated resource scheduling and avoids the security issues of public networks. However, it requires a significant investment to purchase and maintain the machines.

A special form of co-located integration is integration as a quantum module of the Modular Supercomputing Architecture (MSA)~\cite{literature_res_109}. Developed at the Jülich Supercomputing Centre (JSC), this architecture does not use homogeneous compute nodes with different accelerators. Instead, it pools similar resources in modules that are connected by an interconnect. Quantum computers fit naturally into this concept, as they can be seen as another computing module that is good at solving specific tasks. Quantum resources at JSC are made available through the Jülich UNified Infrastructure for Quantum computing (JUNIQ)~\cite{juniq_1,juniq_2}, which provides access via a platform-as-a-service model.

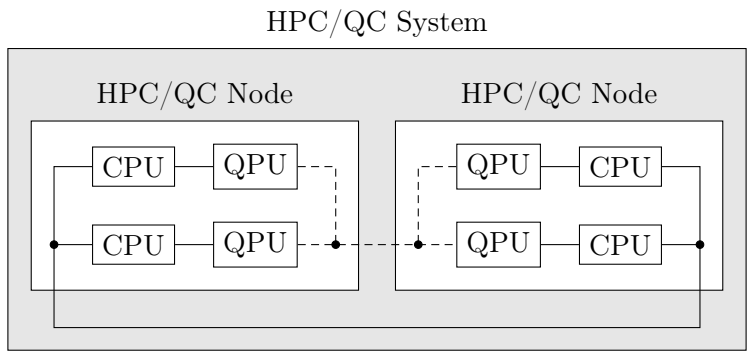
\begin{figure}[htp]
    \centering
    \begin{tikzpicture}
\node(c1) [draw, rectangle, fill=white] {CPU};
\node(q1) [draw, rectangle, fill=white, right=5mm of c1] {QPU};
\node(c2) [draw, rectangle, fill=white, below=5mm of c1] {CPU};
\node(q2) [draw, rectangle, fill=white, right=5mm of c2] {QPU};

\node(q3) [draw, rectangle, fill=white, right=3.7cm of c1] {QPU};
\node(c3) [draw, rectangle, fill=white, right=5mm of q3] {CPU};
\node(q4) [draw, rectangle, fill=white, right=3.7cm of c2] {QPU};
\node(c4) [draw, rectangle, fill=white, right=5mm of q4] {CPU};

\draw (c1) -- (q1);
\draw (c2) -- (q2);
\draw (c3) -- (q3);
\draw (c4) -- (q4);

\coordinate(c2l) at ($(c2.west) - (0.5,0)$);
\coordinate(c4l) at ($(c4.east) + (0.5,0)$);
\filldraw[black] (c2l) circle (0.5mm);
\filldraw[black] (c4l) circle (0.5mm);
\draw (c1) -| (c2l) -- ++(0,-1.1) -| (c4l) |- (c3);
\draw (c2) -- (c2l);
\draw (c4) -- (c4l);

\coordinate(q2r) at ($(q2.east) + (0.5,0)$);
\coordinate(q4r) at ($(q4.west) - (0.5,0)$);
\filldraw[black] (q2r) circle (0.5mm);
\filldraw[black] (q4r) circle (0.5mm);

\draw[densely dashed] (q2) -- (q4);
\draw[densely dashed] (q1) -| (q2r);
\draw[densely dashed] (q3) -| (q4r);

\begin{pgfonlayer}{middle}
    \node (n1) [draw, fill=white, fit={(c1) (q1) (c2l) (q2r) (c2) (q2)}, inner sep=3mm] {};
\end{pgfonlayer}
\node(t1)[above=0mm of n1.north] {HPC/QC Node};
\begin{pgfonlayer}{middle}
    \node (n2) [draw, fill=white, fit={(c3) (q3) (c4l) (q4r) (c4) (q4)}, inner sep=3mm] {};
\end{pgfonlayer}

\node(t2)[above=0mm of n2.north] {HPC/QC Node};
\begin{pgfonlayer}{background}
    \node(s) [draw, fill=lightgray, fit={(n1) (n2) (t1) (t2) ($(c2l) - (0, 1.1)$)}, inner sep=3mm] {};
\end{pgfonlayer}
\node (st) [above=0mm of s.north] {HPC/QC System};
\end{tikzpicture}
    \caption{Schematic of the tight or on-node integration model. The HPC node consists of CPUs and QPUs that are directly connected. The QPUs can have an optional quantum interconnect (dashed lines), even across different nodes.}
    \label{fig:macro_architecture_tight}
\end{figure}

Since quantum computers are bulky and require an isolated environment, today's systems are loosely integrated. Advances in hardware development may allow tighter integration in the future. A quantum computer could be integrated in a similar way to other accelerators such as GPUs at the node level. This integration could be anything from being connected to the node's motherboard to being on the same die as the CPU. This architecture would bring reduced latency and also capabilities such as direct access to main memory. Figure~\ref{fig:macro_architecture_tight} shows a schematic of this architecture. The HPC system consists of nodes consisting of CPUs and integrated QPUs. We use the term QPU rather than quantum computer here because the CPU could act as a QCP and the quantum hardware is not an independent system in this case (see Figure~\ref{fig:definitions}). While it is not yet possible to build such a system, Reference~\cite{literature_res_140} proposes the concept of a virtual quantum processor. The processor's components are classically simulated and can be replaced with real hardware as soon as it becomes available.

The benefits of tighter integration are assessed in Reference~\cite{literature_res_110} for the Quantum Approximate Optimization Algorithm~\cite{qaoa_original,qaoa_review}, which involves a great amount of interaction between classical and quantum hardware. The authors demonstrate that execution time is significantly reduced when the quantum computer is transitioned from a cloud with 50~ms latency to a local network with 1~ms latency. However, there is little further improvement if the quantum computer is integrated into the same system on a chip with an assumed latency of 25~\textmu s. This improvement depends on the circuit execution time, though. The longer the circuit takes to execute, the less relevant the communication time becomes. Another benefit of on-premises integration that is not taken into account is that scheduling classical and quantum resources together could reduce waiting time on the quantum computer. This topic is discussed further in Section~\ref{sec:middleware}.

Quantum accelerator density does not play a major role today because quantum computers are experimental machines and it is already costly for an HPC site to host a single quantum computer. In the future, however, it would be important to avoid oversubscription of quantum computers to the point where the speed advantage is negated by excessively long waiting times~\cite{literature_res_458}. Reference~\cite{literature_res_199} mentions an additional possible benefit of higher quantum accelerator density. If it is possible to connect the accelerators by quantum interconnects, the computational space is $2^{Nn}$ instead of $N2^n$, where $N$ is the number of accelerators and $n$ is the number of qubits per accelerator. Thus, additional accelerators exponentially increase the computational space. The use of multiple quantum computers for a common computation is called distributed quantum computing~\cite{dqc} and is a separate field of research that we do not address here.

\subsection{Micro-Architecture}

\begin{figure}[htp]
    \centering
    \begin{tikzpicture}
  \draw (0,0) rectangle ++(1,0.7);
  \node at (0.5,0.35) {CPU};
  \draw (2,0) rectangle ++(1,0.7);
  \node at (2.5,0.35) {QCP};
  \draw (4,0) rectangle ++(1,0.7);
  \node at (4.5,0.35) {PEL};
  \draw (6,0) rectangle ++(1,0.7);
  \node at (6.5,0.35) {MCE};
  \draw (8,0) rectangle ++(1.5,0.7);
  \node at (8.75,0.35) {Qubits};
  \draw (0,-1.7) rectangle ++(7,0.7);
  \node at (3.5,-1.35) {Synchronization Clock};
  \draw (0.5,-1)[-angle 45] -- ++(0,1);
  \draw (2.5,-1)[-angle 45] -- ++(0,1);
  \draw (4.5,-1)[-angle 45] -- ++(0,1);
  \draw (6.5,-1)[-angle 45] -- ++(0,1);
  \draw (1,0.35)[angle 45-angle 45] -- ++(1,0);
  \draw (3,0.35)[angle 45-angle 45] -- ++(1,0);
  \draw (5,0.35)[angle 45-angle 45] -- ++(1,0);
  \draw (7,0.35)[angle 45-angle 45] -- ++(1,0);
  \draw (0.5,1.7) rectangle ++(2.5,0.7);
  \node at (1.75,2.05) {Arbiter};
  \draw (0,3.4) rectangle ++(3,0.7);
  \node at (1.5,3.75) {Main Memory};
  \draw (0.25,0.7)[angle 45-angle 45] -- ++(0,2.7);
  \draw (0.75,0.7)[angle 45-] -- ++(0,1);
  \draw (2.5,0.7)[angle 45-] -- ++(0,1);
  \draw (1.75,2.4)[angle 45-] -- ++(0,1);
\end{tikzpicture}
    \caption{Overview of the microarchitecture of a hybrid quantum-classical processor. It is a simplified version of the overview from Reference~\cite{literature_res_214}. The main memory contains instructions that are forwarded to either the CPU or the QCP. The QCP performs operations on the qubits via the physical execution layer (PEL), making the QCP technology-independent.}
    \label{fig:micro_architecture}
\end{figure}
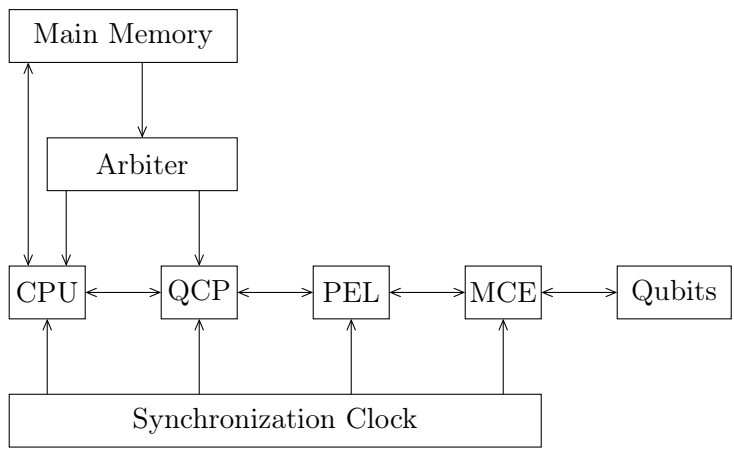

In the last section, we described a high-level architecture in which the quantum computer is a single block. However, there are publications that outline a much more fine-grained architecture. Reference~\cite{literature_res_214} sketches an architecture for a classical host processor combined with a quantum accelerator. This system aligns well with the tight on-node integration model. Figure~\ref{fig:micro_architecture} shows a simplified block diagram of this architecture. We use different names for some of the blocks than those in the original publication to be consistent throughout this paper. A detailed illustration of this architecture is shown in the cited paper.

The main principle is that the main memory contains the data and instructions for the classical as well as for the quantum processor. An arbiter forwards instructions to either the classical CPU or the QCP, which decodes them and uses the MCE to create the appropriate waveforms to control the qubits. The authors introduce an additional layer, the physical execution layer (PEL), which makes the QCP independent of the technology used. A clock is connected to all stages for synchronization, which is important for error correction. The CPU and QCP can communicate through a register file.

A similar architecture is proposed in Reference~\cite{literature_res_480}; however, the focus there is on the QCP architecture. Classical instructions are processed by a simple RISC-V processor. Instructions for the quantum computer are in a format similar to eQASM~\cite{eqasm} but with improved addressing that allows for more qubits. Additionally, the QCP has a histogram unit that accumulates measurement results in hardware and continuously sorts them. This enables the transmission of only the highest probabilities to the host processor, reducing communication and computational effort. The proposed QCP architecture is validated on a field programmable gate array.

\section{Software Stack}
\label{sec:software}

A software stack is a set of tools that work together to run applications on a given piece of hardware. These tools typically form hierarchical levels that build on each other, with the application at the top and the hardware at the bottom. The specific components depend on the system and often include firmware, an operating system, middleware with communication layers, and various APIs on which the application is based. In the HPC context, we must differentiate between the compile-time and runtime tool chains.

There are separate software stacks for quantum computing and HPC. Reference~\cite{literature_res_99} compares typical quantum and HPC software stacks. The authors state that HPC stacks contain compilers, runtimes, and libraries that aim to reduce execution time by doing as much as possible at compile time. Conversely, typical quantum programs are usually written in Python, which is interpreted at runtime. This is partly due to the convenience of using Python and partly because the structure of quantum programs is not fixed and can change during runtime. The authors argue that, for a hybrid software stack, the runtime components of quantum computing must be integrated into the HPC stack. Additionally, software environments, such as monitoring and scheduling, should be unified.

The earliest paper we found about a hybrid software stack in our research is from 2016~\cite{literature_res_211}, in which the authors describe that existing papers refer to either the instruction set architecture or the programming model, i.e. layers at the bottom or top of the software stack. They note a gap in the literature regarding middleware, such as communication protocols. In subsequent years, more papers have discussed the hybrid quantum-classical software stack as a whole~\cite{literature_res_34,literature_res_244,literature_res_313,literature_res_375,literature_res_476,literature_res_479,literature_res_481,literature_res_485}.

We identify the main points shown in Figure~\ref{fig:sw_hw_stack} that must be addressed in order to establish a common stack. A hybrid application is based on a hybrid programming model. At runtime, the hybrid middleware is responsible for handling communication between classical and quantum processors, as well as scheduling. Within the loose integration model, we have separate classical and quantum hardware with respective firmware running. Within the tight integration model, hybrid hardware with a common firmware could also exist in the future.

At compile time, a hybrid compiler creates a hybrid executable. However, we define the term ``hybrid compiler'' very broadly. It could be a classical compiler and a quantum compiler that are executed separately for the respective parts of an application. Alternatively, it could be a compiler with a common instruction set for classical and quantum code. Compilation can also occur during runtime (i.e., just-in-time compilation). This is much more common for quantum circuits than for classical HPC code since parameterized circuits can be optimized more efficiently when the values of the parameters are known. Reference~\cite{literature_res_95} outlines the differences between classical and quantum compilers. Because quantum computer programming is sometimes more akin to hardware description, it is difficult to standardize the compilation stacks.

\begin{figure}[htp]
    \centering
    \begin{tikzpicture}
\draw[black] (0,0) rectangle ++(8,4.8);
\draw[black] (0,0.8) -- ++(8,0);
\draw[black] (0,1.6) -- ++(8,0);
\draw[black] (0,3.2) -- ++(8,0);
\draw[black] (0,4) -- ++(8,0);
\draw[black,dashed] (4,0) -- ++(0,1.6);
\node at (2,0.4) {Classical Hardware};
\node at (6,0.4) {Quantum Hardware};
\node at (2,1.2) {Classical Firmware};
\node at (6,1.2) {Quantum Firmware};
\node at (4,2.4) {Hybrid Middleware};
\node at (4,3.6) {Hybrid Programming Model};
\node at (4,4.4) {Hybrid Application};
\end{tikzpicture}
    \caption{Software stack for hybrid quantum-classical computing. A hybrid application is based on a hybrid programming model. During runtime, the hybrid middleware manages the communication and scheduling of the classical and quantum hardware, which are managed by their respective firmware. In the tight integration model, we can think of a hybrid firmware that manages the hybrid hardware.}
    \label{fig:sw_hw_stack}
\end{figure}

\subsection{Applications}
\label{sec:applications}

Although we do not focus directly on applications, we nevertheless list here the publications that we found during our literature research dealing with hybrid quantum-classical applications. On the one hand, these applications motivate the development of hybrid systems; without them, one might question the need for such systems. On the other hand, we can also derive requirements for systems from applications. For instance, the frequency with which classical and quantum systems communicate in an application can be used to deduce the importance of latency for communication. However, the applications mentioned here should be viewed as examples, not as a complete list. An extensive collection of quantum algorithms, some of which are hybrid, can be found in Reference~\cite{literature_res_17}, for example. A review of hybrid algorithms in quantum chemistry can be found in Reference~\cite{literature_res_41}. A hybrid end-to-end simulation of a chemical system is presented in Reference~\cite{literature_res_294}. References~\cite{literature_res_78,literature_res_87} present ideas for applying quantum computing to power grid simulations.

In practice, every quantum application requires classical resources to run. A circuit is typically created or adapted in a pre-processing step to solve a specific problem. A post-processing step follows to interpret or visualize the results. This step may include error mitigation, which is particularly relevant in the noisy intermediate-scale quantum (NISQ) era~\cite{nisq_era}. However, these classical resources do not have to be HPC systems. Often, a user's personal computer is sufficient. For fault-tolerant quantum computing, classical feedback loops are required for error correction. These classical computations, however, are rather simple but must be performed with low latency. Therefore, they will be executed on control hardware close to the QPU rather than on an HPC system. A definition of hybrid algorithms can be found in Reference~\cite{literature_res_469}, where the authors define a hybrid algorithm as one that requires ``nontrivial amounts of both quantum and classical computational resources to run.''  With ``nontrivial'', they exclude simple classical feedback loops.

Reference~\cite{literature_res_32} argues that we need a ``killer application'' that enables useful computations in the NISQ era, naming variational quantum algorithms as potential candidates. In variational algorithms, we try to find the lowest expectation value of a problem Hamiltonian with a parametrized ansatz implemented on a quantum computer. The parameters of the ansatz are optimized using a classical optimizer. Thus, variational algorithms are hybrid. They are characterized by frequent switching between classical and quantum hardware because the quantum circuit must be executed again after each parameter adjustment. There are countless works on variational algorithms. In our literature review, we came across the references~\cite{literature_res_18,literature_res_245,literature_res_279,literature_res_311,literature_res_280,literature_res_456,literature_res_477,literature_res_13,literature_res_366}. If problems are too large to be executed on a specific device, it is possible to break them down into sub-problems and solve them on the quantum computer in order to solve the overall problem with the partial results. Reference~\cite{literature_res_133} describes such a procedure for combinatorial optimization problems that are solved on gate-based quantum computers and quantum annealers.

Another common type of hybrid application in our research is quantum machine learning (QML) algorithms. In QML, classical machine learning algorithms or neural networks, are implemented on quantum computers~\cite{qml_overview}. For neural networks, the rotation angles of parameterized quantum circuits are trainable parameters. These algorithms are often hybrid because not all layers run on the quantum computer. Although the advantage of QML algorithms over classical machine learning remains unclear, many of the references in our literature research address this topic~\cite{literature_res_12,literature_res_61,literature_res_134,literature_res_482,literature_res_267,literature_res_316,literature_res_337,literature_res_349}.

Quantum annealers, also known as analog quantum computers, are specific types of quantum computers designed to solve quadratic optimization problems. More precisely, they can solve problems that can be formulated as quadratic unconstrained binary optimization. Like classical analog computers, quantum annealers face the issue that problems must fit into the topology of the processor. For problems that exceed the processor's capacity, hybrid solvers are used, in which subproblems are solved on the annealer to find the solution to a larger optimization problem. In this sense, the algorithms are hybrid. References~\cite{literature_res_14,literature_res_73,literature_res_207,literature_res_205,literature_res_190} use a quantum annealer in a hybrid setting to solve optimization problems.

\subsection{Programming Models and Frameworks}
\label{sec:programming_models}
There are a large number of quantum programming languages or quantum extensions to existing programming languages~\cite{qc_lang_overview_2020,literature_res_55}. In this section, we discuss programming frameworks and models that are specifically designed for hybrid quantum-classical setups. Ideally, such a framework would be extensible and modular, as well as hardware-agnostic, in order to cover the wide variety of hardware technologies available today. It should also be compatible with existing HPC tools~\cite{literature_res_126}. An extensive review of quantum programming tools can be found in Reference~\cite{literature_res_372}, in which the authors evaluate existing tools based on their HPC integration capabilities. They conclude that existing tools only partially cover this integration and recommend that future tools prioritize HPC integration. The authors also note the lack of benchmarks for hybrid systems. Since the paper is from 2023, it does not cover the latest frameworks (see Figure~\ref{fig:overview_years}; many publications about hybrid quantum-classical computing have appeared in 2024).

There is a similar publication that reviews tools for analog quantum computing~\cite{literature_res_256}. The authors note that these tools are not ready for integration into HPC systems, with the exception of the XACC framework (see Section~\ref{sec:xacc}). This is consistent with our findings from this literature review. Although we did not explicitly search for gate-based quantum computers, almost all publications deal with this type of quantum computer rather than analog quantum computers. This is likely due to the greater popularity and prevalence of gate-based systems, as well as their greater similarity to classical digital computers. In the following, we take a closer look at different categories of programming models.

\subsubsection{Workflow Models}
\label{sec:workflow_models}
Workflow models have their origins in the automation of business processes. The Workflow Management Coalition defines a workflow as ``the computerised facilitation or automation of a business process, in whole or part''~\cite{workflow_coalition}. They mention that workflows can be performed manually, but in practice are most often automated by IT systems. Workflows allow loosely coupled and distributed components to work together. Remote services can be invoked as a black box without requiring knowledge of the internals~\cite{workflow_scientific}. The concept has been adapted for scientific workflows to process scientific data. There are several standards for workflow models that can be executed by workflow engines~\cite{workflow_model_1,workflow_model_2,workflow_model_3,workflow_model_4}. Workflow models allow the integration of new types of computing resources in a natural way. They fit well into the loosely coupled client-server model. For this reason, several publications deal with the integration of quantum computers into workflows, which are presented below. A general description of such a setup, along with the necessary steps for modeling the tasks and dependencies, can be found in Reference~\cite{literature_res_26}. Table~\ref{tab:workflow_models} provides an overview of the models presented. A drawback of workflow models is the high latency of the individual services, which is particularly noticeable when switching frequently between classical and quantum hardware.

Reference~\cite{literature_res_459} outlines a general concept for hybrid quantum-classical workflows. The authors define a set of classical tasks, $T$, and a set of quantum candidates, $T' \subset T$. For each task in $T'$, there is an equivalent quantum task in $Q$, and a function $f: T \mapsto Q$ defines how the classical task is mapped to its quantum equivalent. For each task in $T'$, a decision node exists in the set $D$ to determine whether the task will be executed on a classical or quantum resource. One possible metric for this decision is the available quantum resources. While this allows for a flexible workflow with a backup solution in case a quantum resource is unavailable, it requires additional effort since classical tasks and their mapping to equivalent quantum tasks must be specified. This paper is conceptual and does not provide an implementation of the outlined concept.

A more applied approach is taken in Reference~\cite{literature_res_471}, which introduces QuantME, an extension to existing workflow languages. The idea is to create a QuantME workflow model that allows the execution of quantum circuits, but also pre- and post-processing steps. This model is then translated into a native workflow model that can be used in existing workflow engines. This extension is prototypically implemented for the Business Process Model and Notation~\cite{bpmn}. For a more detailed example that includes automatic provisioning of execution environments, see Reference~\cite{literature_res_466}.

To avoid the disadvantage of high queuing times with fine-grained dependencies, such as those found in variational algorithms, Reference~\cite{literature_res_484} proposes the use of hybrid runtimes. Instead of running the classical part on a classical runtime and the quantum part on a quantum runtime, both parts are run together on a hybrid runtime, where classical and quantum resources are more tightly coupled to reduce latencies. The paper introduces a framework called MODULO to automatically find such hybrid parts in QuantME workflows and rewrite them for hybrid backends. Although QuantME provides a framework for modeling workflows, each workflow's individual steps must be implemented separately. To avoid repeatedly implementing the various steps required for variational quantum algorithms, Reference~\cite{literature_res_467} offers a set of commonly needed services that can be readily used.

Another dataflow framework is Tierkreis~\cite{literature_res_468}. It provides a strong static type system that allows static code analysis to detect inconsistencies in the code before it is deployed on expensive quantum resources. The framework provides some primitive operations for control flow and data processing, but also allows calls to external functions via gRPC~\cite{grpc}. An open source Python package of the framework is available on Github~\cite{tierkreis_repo}.

Reference~\cite{literature_res_370} uses a workflow model to access remote quantum resources from the Oak Ridge Leadership Computing Facility (OLCF). The paper is a good illustration of how the remote integration model can be implemented via workflows. Users connect to the HPC system via a terminal. The HPC system is used to generate quantum circuits and corresponding inputs, and then connects to a remote quantum resource from Rigetti, IBMQ, or Quantinuum. The results are then returned to the HPC system for post-processing. This setup allows easy use of different remote quantum computers, but is not suitable for a large number of interactions between quantum and classical hardware, since the queues of remote machines can be a bottleneck.

The QCCP model~\cite{literature_res_2} is specifically designed for hybrid quantum-classical programming in Python. It allows functions to be annotated with \texttt{@qccp.ctask} and \texttt{@qccp.qtask} decorators to define classical and quantum tasks. As with other models, a direct acyclic graph (DAG) is created from the task dependencies. A global workflow scheduler keeps track of these dependencies and sends tasks whose dependencies are fulfilled to the respective classical and quantum backends. The backends are managed by local schedulers.

Qonductor takes a similar approach~\cite{literature_res_309}. Rather than using Python annotations, this approach uses a Python library. This library contains functions for creating and deploying workflows. The API provides frequently used algorithms, as well as pre- and post-processing options, such as circuit cutting and error mitigation. A workflow is created using these functions and a YAML configuration file that specifies the necessary hardware. The workflow is deployed on a leader node, which executes the workflow's individual steps on the classic and quantum resources. The scheduling approach of Qonductor is described in Section~\ref{sec:middleware}.

\begin{table}[ht]
    \centering
    \caption{Workflow models for hybrid quantum-classical computing with the corresponding references.}
    \begin{tabular}{l l l}
        \hline
        Model & Description & References \\\hline\hline
        QuantME &Quantum extension for existing workflow models.& \cite{literature_res_471,literature_res_466,literature_res_484,literature_res_467}\\\hline
        Tierkreis & Dataflow framework with a strong static type system. & \cite{literature_res_468,tierkreis_repo}\\\hline
        Parsl &Framework for accessing quantum resources at the OLCF.&\cite{literature_res_370}\\\hline
        QCCP &Framework that uses Python decorators to model workflows.&\cite{literature_res_2}\\\hline
        Qonductor & Library based framework to model and execute hybrid workflows. & \cite{literature_res_309}\\\hline
    \end{tabular}
    \label{tab:workflow_models}
\end{table}

\subsubsection{Function as a Service}
Function as a service (FaaS) is a programming paradigm stemming from cloud computing. It implements a serverless execution model. Serverless does not refer to the absence of physical servers, but rather to the abstraction of those servers from the user. Because it comes from cloud computing, FaaS is well-suited to the remote integration model, in which cloud-based quantum resources are used.

Reference~\cite{literature_res_465} introduces QFaaS, a quantum FaaS framework. It provides access to quantum resources from different vendors and combines them with state-of-the-art cloud computing methods. Quantum software engineers can write quantum functions and deploy them to the system. There are templates available for commonly used quantum SDKs, such as Qiskit~\cite{qiskit}. The function itself is written in Python. After deployment, regular users can use the existing functions to create hybrid quantum-classical workflows that can be executed on different backends, including simulators and real devices from commercial vendors. The QFaaS code~\cite{qfaas_github} and a deployment guide~\cite{qfaas_deployment} are available online.

Kernel as a service (KaaS)~\cite{literature_res_75} is inspired by FaaS. It is a general model for heterogeneous hardware accelerators. However, the paper also demonstrates an implementation of this concept on quantum computers. In the KaaS model, applications are composed as collections of kernels. Rather than executing these kernels directly on accelerators, they are registered and invoked on a KaaS server, which then assigns them to available computing resources. This allows for the sharing of resources across kernels, which can increase hardware utilization.

\subsubsection{Extensions of HPC Frameworks}
\label{sec:hpc_models}
Over the past few decades, HPC systems have become more and more parallel and heterogeneous. Today's systems use accelerators, primarily GPUs, and programming models and software frameworks have been developed to simplify programming for these systems. Rather than developing entirely new models, it makes sense to adapt existing models for use with quantum computers, enabling users to remain in their familiar environments. We discuss multiple such frameworks below. Table~\ref{tab:hpc_models} provides an overview of the frameworks.

C++ is the most widely used programming language in high-performance computing (HPC). Thus, references \cite{literature_res_470} and \cite{literature_res_373} offer a direct extension of C++ for quantum computers. Both publications achieve this by extending the Clang compiler~\cite{clang} to enable the compilation for hybrid architectures from a single source. Reference~\cite{literature_res_470} defines an instruction set architecture for quantum accelerators. Quantum instructions are marked with the \texttt{quantum\_kernel} attribute. These instructions must be resolved at compile time to prevent costly quantum compilation at run time. The only exception are parameters for quantum gates, which can be annotated with \texttt{quantum\_shared\_var} and are important for variational algorithms, for example. A library provides commonly used quantum gates. During compilation, classical and quantum sources are separated and sent to their respective compilers. Then, they are linked to a common, extensible and linkable file for quantum (ELFQ), which is an extension of the standard ELF format. Reference~\cite{literature_res_373}, on the other hand, uses pragma directives rather than specific attributes to mark parts of the code that run on the quantum computer.

The quantum computer extensions for OpenMP~\cite{literature_res_363} and OmpSs-2~\cite{literature_res_246} are also based on special pragma directives. OpenMP is one of the most widely used programming models in high-performance computing (HPC)~\cite{openmp} for parallel programming of shared memory systems. The \texttt{\#pragma omp target} directive is used to offload parts of the code to a quantum computer. Offloading is synchronous by default, but it can be made asynchronous with the \texttt{nowait} clause. OmpSs-2 is similar to OpenMP in terms of syntax, but implements a task-based programming model in which parallelism is modeled implicitly by task dependencies~\cite{ompss2}. These tasks are functions that are annotated with the \texttt{\#pragma omp task} directive and can be offloaded to a quantum computer with the \texttt{device(qpu)} clause. In this case, the annotated function is not implemented in the classical code; rather, an OpenQASM file~\cite{openqasm} containing the implementation is sent to the quantum computer during runtime.

OpenCL~\cite{opencl} and CUDA~\cite{cuda} are two frameworks used to program GPUs. Since quantum computers are considered accelerators and are often compared to GPUs, adapting these frameworks for quantum computers is a logical next step. Both frameworks allow for the implementation of kernels that are executed on the accelerator. A subset of the C99 standard~\cite{c99} must be used to write the kernel for OpenCL. In the quantum computing extension of OpenCL~\cite{literature_res_50}, the kernel describes a circuit at the gate level using a syntax similar to the C implementation of Qulacs~\cite{qulacs}. The OpenCL implementation for quantum computers, along with examples, can be found in Reference~\cite{opencl_implementation}. CUDA Quantum~\cite{literature_res_79} also allows users to define kernels at the gate level and compile them into QIR~\cite{literature_res_464}. This code can be executed on either a real device or a simulation backend. One simulator is cuQuantum~\cite{cuquantum}, which uses GPUs for simulation.

The Quingo framework~\cite{literature_res_121} is inspired by OpenCL as well and is based on a refined heterogeneous quantum-classical computation model. This model builds upon the QRAM model introduced by Knill~\cite{qram_model} by accounting for parts of the classical code that require low-latency real-time execution. This is important for error correction, for example. This model fits well with Figure~\ref{fig:definitions} if we assume that a classical computer is connected to the quantum computer via the classical interconnect. Users can write programs for the QCP that interact with the QPU via very low latency, as well as programs for the classical computer that interact with higher latency but can be more computationally intensive.

Users must decide which parts of their programs should run on the QCP and which parts should run on the classical computer connected to the quantum computer. The computationally intensive part is written in a classical programming language, such as C++ or Python. The QCP and quantum operations are written in the programming language Quingo. The framework provides a compiler that translates the code into eQASM~\cite{eqasm}, a quantum assembly language. Examples are available on GitHub~\cite{quingo_repo}.

\begin{table}[ht]
    \centering
    \caption{Overview of programming models and frameworks that are based on existing HPC models.}
    \begin{tabular}{l l l}
        \hline
        Name / Description & Extends / is based on & References \\\hline\hline
        LLVM for Quantum & C++, Clang & \cite{literature_res_470}\\\hline
        Pragma Based C++ Framework for Quantum & C++, Clang & \cite{literature_res_373}\\\hline
        Quantum Extension for OpenMP & OpenMP & \cite{literature_res_363}\\\hline
        Quantum Extension for OmpSs-2 & OmpSs-2 & \cite{literature_res_246}\\\hline
        Quantum Extension for OpenCL & OpenCL & \cite{literature_res_50}\\\hline
        CUDA Quantum & CUDA & \cite{literature_res_79}\\\hline
        Quingo & OpenCL & \cite{literature_res_121}\\\hline
    \end{tabular}
    \label{tab:hpc_models}
\end{table}

\subsubsection{Extensions of Quantum Computing Frameworks}
\label{sec:qc_models}

A large number of programming models and frameworks have been developed specifically for quantum computers. For example, most hardware manufacturers offer their own programming frameworks for their hardware. Most of these frameworks provide a Python API and do not focus on HPC. Here, we present frameworks that focus specifically on integrating with HPC or that have been extended to integrate quantum computing with HPC.

PennyLane is the framework of Xanadu, a manufacturer of photonic quantum computers~\cite{literature_res_461,literature_res_460}. However, the framework is not limited to photonic computing. It implements the concept of differentiable programming of quantum computers. This concept involves calculating the derivative of a parametrized quantum circuit. It can be used for variational quantum algorithms and quantum machine learning, for example. The Python-based interface is compatible with machine learning libraries such as PyTorch~\cite{pytorch} and TensorFlow~\cite{tensorflow}, which have efficient CPU and GPU implementations.

Amazon AWS offers cloud-based computing resources~\cite{amazon_aws}. In addition to classical computing resources, users can access quantum computers from various vendors. These devices can be programmed using the Python-based Amazon Braket framework~\cite{amazon_braket}. Since classical and quantum resources are available in the cloud, AWS also offers executing hybrid jobs on cloud resources~\cite{literature_res_472}. Hybrid jobs have prioritized access to quantum hardware so that the queue for accessing the quantum hardware does not stall the classical resources waiting for quantum results. With IBM's Qiskit~\cite{qiskit}, it is also possible to perform classical calculations close to the quantum computer, which can bring significant speedup when executing variational algorithms, for example~\cite{literature_res_137}. 

\subsubsection{XACC and QCOR}
\label{sec:xacc}

The eXtreme-scale ACCelerator (XACC) framework was designed from the beginning as a hybrid quantum-classical framework. Introduced in 2018~\cite{literature_res_195}, it was described in more detail in a subsequent publication~\cite{literature_res_195} and extended for the use with QIR~\cite{literature_res_462,literature_res_464}. Additionally, there are examples for the use of XACC in the domains of nuclear physics~\cite{literature_res_475} and chemistry simulations~\cite{literature_res_69}. The framework defines a platform model consisting of three parts. A host (CPU) that communicates with a quantum computer via an accelerator buffer. The host creates the accelerator buffer, which stores the results from the quantum computer.

As with some of the frameworks in Section~\ref{sec:hpc_models}, users can write quantum kernels embedded in their C++ code. This can be done using the custom XACC assembly language (XASM) or other commonly used languages. Quantum kernels are first translated into an intermediate representation, then compiled for a specific target, which may be a simulator or a real device. A simulator based on tensor networks has been specially developed for use with XACC for the simulation of large circuits~\cite{literature_res_463}. Unlike most other frameworks, XACC also supports analog quantum computers. Listing~\ref{lst:xacc_kernel_example} shows an example kernel from~\cite{literature_res_158}. The kernel function is annotated with \texttt{\_\_qpu\_\_}. The first parameter of the function is always the aforementioned accelerator buffer. In this example, the function is implemented in XASM, which describes the quantum program at the gate level.

\begin{listing}[ht]
\begin{lstlisting}[
    label={lst:xacc_kernel_example},
    language=C++,
    caption={XACC kernel example~\cite{literature_res_158}. In this example, there are two kernels, annotated with \texttt{\_\_qpu\_\_} and implemented in Quil~\cite{quil}. The kernels get translated to OpenQASM to run on IBMQ devices.}
    ]
auto qpu = xacc::getAccelerator("ibm:back-end");
auto quil = xacc::getCompiler("quil");
auto ir = quil->compile(R"(
__qpu__ void ansatz(AcceleratorBuffer q, double x) {
    X 0
    Ry(x) 1
    CX 1 0
}
__qpu__ void X0X1(AcceleratorBuffer q, double x) {
    ansatz(q, x)
    H 0
    H 1
    MEASURE 0 [0]
    MEASURE 1 [1]
}
)", qpu);
auto x0x1 = ir->getComposite("X0X1");
auto openqasm = xacc::getCompiler("openqasm");
auto oqasmSrc = openqasm->translate(x0x1);
\end{lstlisting}
\end{listing}

Closely related to the XACC framework is the QCOR specification~\cite{literature_res_156}. It specifies, how languages can be extended to enable single-source quantum-classical programming by defining a memory model and data structures. The memory model defines memory spaces on the host and quantum device. Memory on the host device can be explicitly allocated. The device memory is allocated implicitly by the compiler if for example host memory is reserved for the results of the quantum device. Memory transfers between host and device memory are also handled by the compiler.

The following data structures are defined:
\begin{itemize}
    \item Observable: Represents rules about how one or more qubits are measured after execution of the circuit.
    \item ObservableTransform: Represents a transformation from one Observable to another Observable.
    \item ResultBuffer: Provides a container to store the measured results of a quantum execution.
    \item ObjectiveFunction: Represents a parametrized scalar function, that can be evaluated for an array of scalar parameters.
    \item Optimizer: Takes an ObjectiveFunction as input and calculates the optimal function parameters.
\end{itemize}

There is an implementation of the QCOR specification for C++~\cite{literature_res_129}. The implementation introduces a compiler called qcor (note the lower case letters in comparison to the upper case letters of the specification name). The qcor compiler provides a plugin for the Clang compiler~\cite{clang} and utilizes the XACC framework to compile quantum resources. The compiler takes care of quantum specific issues like placement and gate level optimization and can be called with similar options like classical compilers. There is an extension to qcor, which allows multithreading~\cite{literature_res_84}. To achieve this, the authors add thread safety to the qcor implementation. They also remove data races that prevent efficient parallelization. Standard C++ constructs such as \texttt{std::thread} or \texttt{std::async} can be used for the parallel code. 

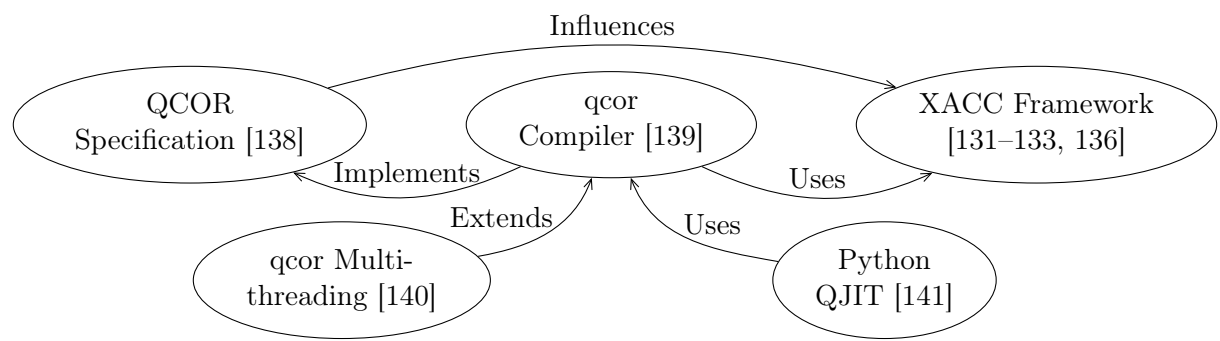
\begin{figure}[htp]
    \centering
    \begin{tikzpicture}[node distance=0.5cm and 1.3cm]
    
    \node[draw, ellipse, align=center] (qcorspec) {QCOR\\Specification~\cite{literature_res_156}};
    \node[draw, ellipse, align=center] (qcorcompiler) [right=of qcorspec] {qcor\\Compiler~\cite{literature_res_129}};
    \node[draw, ellipse, align=center] (xacc) [right=of qcorcompiler] {XACC Framework\\\cite{literature_res_158,literature_res_195,literature_res_462,literature_res_475}};
    \node[draw, ellipse, align=center] (multithread) [below=of qcorspec, xshift=2cm] {qcor Multi-\\threading~\cite{literature_res_84}};
    \node[draw, ellipse, align=center] (python) [below=of xacc, xshift=-2cm] {Python\\QJIT~\cite{literature_res_91}};
    
    \path (qcorcompiler) edge[-angle 45, in=335, out=205] node[above] {Implements} (qcorspec);
    \path (qcorcompiler) edge[-angle 45, in=205, out=335] node[above] {Uses} (xacc);
    \path (qcorspec) edge[-angle 45, in=165, out=15] node[above] {Influences} (xacc);
    \path (multithread) edge[-angle 45, in=250, out=10] node[left, yshift=2mm, xshift=2mm] {Extends} (qcorcompiler);
    \path (python) edge[-angle 45, in=290, out=170] node[right, yshift=2mm, xshift=-2mm] {Uses} (qcorcompiler);
    
\end{tikzpicture}
    \caption{Interdependencies between QCOR and XACC projects.}
    \label{fig:xacc_qcor}
\end{figure}

While the above work refers to C++ implementations, there is also a language extension for Python to enable hybrid quantum-classical programming~\cite{literature_res_91}. This extension uses the qcor compiler to compile Python functions provided with the \texttt{@qjit} decorator just-in-time for quantum computing. The target audience for this framework is primarily users who want to quickly prototype something. As we can see, there are various projects connected to XACC and QCOR. To make things clearer, Figure~\ref{fig:xacc_qcor} shows the dependencies and associated publications.

\subsection{Middleware}
\label{sec:middleware}

Middleware generally refers to the software layers between the operating system and the application. In the field of HPC, the need for middleware arises from the fact that operating systems, such as Linux, do not inherently provide sufficient support for distributed and parallel computing~\cite{middleware_hpc}. Consequently, HPC middleware primarily supports communication between computing nodes and resource scheduling. Middleware is closely linked to programming models (see Section~\ref{sec:programming_models}). A programming model describes how applications are written, while middleware implements the necessary layers, e.g., for exchanging data to execute applications.

One of the middleware's main tasks is resource scheduling. In the context of HPC, scheduling refers to the assignment of computational jobs to available computing resources. This is typically done at the job level using batch schedulers, such as Slurm~\cite{slurm}. Users specify the number and type of resources required to run a job in job scripts and submit the job to a queue. Once the necessary resources are available, the job runs. Typically, the resources are reserved at the node level for the entire runtime of a job. For instance, if GPUs are not used during a job's runtime, they cannot be used by a different job. For a future tight integration model of quantum computers, we can assume that we can use scheduling models similar to those for GPUs if every node is equipped with a QPU~\cite{literature_res_23}.

However, with the loose integration model, quantum computers become scarce resources. Typically, there are many more classical nodes that offload circuits to them~\cite{literature_res_476}. It would be inefficient to reserve a quantum computer for the entire duration of a job if it is only needed for a fraction of the time. This would prevent other jobs from using the quantum computer. One solution is to use heterogeneous Slurm jobs, as proposed in Reference~\cite{literature_res_357}. This involves splitting jobs across different hardware to avoid blocking the quantum device for the entire runtime of the classical hardware. The NordIQuEst infrastructure~\cite{literature_res_321} employs a setup in which users submit hybrid jobs via Slurm to a classical HPC machine. This machine then accesses the quantum computer via a REST API~\cite{rest_api}. A similar setup is presented in Reference~\cite{literature_res_348}, consisting of two Slurm clusters. The first cluster includes the HPC system. The second cluster is connected to the Quantum Inspire API~\cite{quantum_inspire}. Users can submit jobs to the first cluster, which then submits quantum jobs to the second cluster, where they are scheduled via the Quantum Inspire API.

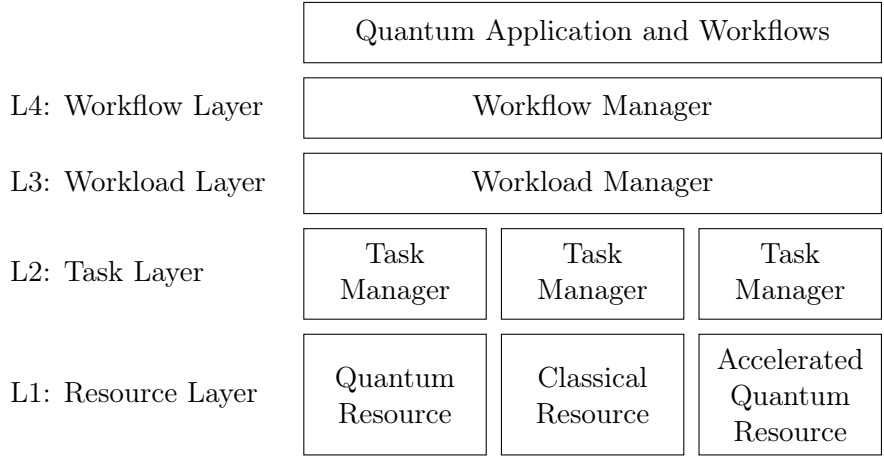
\begin{figure}[htp]
    \centering
    \begin{tikzpicture}
\draw[black] (0,0) rectangle ++(2.4,1.6);
\node[align=center] at (1.2,0.8) {Quantum\\Resource};
\draw[black] (2.6,0) rectangle ++(2.4,1.6);
\node[align=center] at (3.8,0.8) {Classical\\Resource};
\draw[black] (5.2,0) rectangle ++(2.4,1.6);
\node[align=center] at (6.4,0.8) {Accelerated\\Quantum\\Resource};

\draw[black] (0,1.8) rectangle ++(2.4,1.2);
\node[align=center] at (1.2,2.4) {Task\\Manager};
\draw[black] (2.6,1.8) rectangle ++(2.4,1.2);
\node[align=center] at (3.8,2.4) {Task\\Manager};
\draw[black] (5.2,1.8) rectangle ++(2.4,1.2);
\node[align=center] at (6.4,2.4) {Task\\Manager};

\draw[black] (0,3.2) rectangle ++(7.6,0.8);
\node[align=center] at (3.8,3.6) {Workload Manager};

\draw[black] (0,4.2) rectangle ++(7.6,0.8);
\node[align=center] at (3.8,4.6) {Workflow Manager};

\draw[black] (0,5.2) rectangle ++(7.6,0.8);
\node[align=center] at (3.8,5.6) {Quantum Application and Workflows};

\node[anchor=west] at (-4,0.8) {L1: Resource Layer};
\node[anchor=west] at (-4,2.4) {L2: Task Layer};
\node[anchor=west] at (-4,3.6) {L3: Workload Layer};
\node[anchor=west] at (-4,4.6) {L4: Workflow Layer};

\end{tikzpicture}
    \caption{Layers of the conceptual quantum-HPC middleware~\cite{literature_res_83}. The system comprises different resources (L1), which are made available to task managers (L2). The workload manager (L3) assigns tasks to the appropriate resource. The workflow model (L4) defines the dependencies between tasks that form the application.}
    \label{fig:mw_stack}
\end{figure}

A general concept for a hybrid quantum-classical middleware is introduced by Reference~\cite{literature_res_83}. This concept is based on a proposal for middleware of workflow systems~\cite{mw_building_blocks}, which has been adapted for hybrid quantum-classical systems. It has four layers (L1--L4), as shown in Figure~\ref{fig:mw_stack}. The highest level, the workflow layer, models both quantum and classical tasks and their interdependencies. Below the workflow layer is the workload layer, which assigns tasks to resources. These resources can be selected based on task requirements or dependencies. For example, it may be beneficial to assign tightly coupled quantum and classical tasks to nearby quantum and classical resources. The tasks assigned to a given resource are handled on the task layer. This layer is responsible for scheduling, error handling, monitoring, etc., on the system's individual resources, which form the lowest layer.

These layers are implemented in Reference~\cite{literature_res_263}, for which the authors use the concept of pilot jobs. In these jobs, resources are dynamically reserved during runtime rather than statically~\cite{pilot_jobs}. This can lead to higher utilization, especially on heterogeneous systems. Quantum-classical hybrid systems are well-suited for this type of execution. At the heart of the framework is the pilot manager (L3), which starts placeholder jobs via the resource management system (e.g., Slurm) and thus reserves classical and quantum resources. A pilot agent (L2) runs on each reserved resource (L1) and executes individual tasks that, together, form applications (L4). This allows the tasks of several applications to be dynamically distributed to available resources at runtime. Reference~\cite{literature_res_81} also notes the importance of intelligent scheduling. The authors present a prototype implementation of middleware based on various open-source projects~\cite{qiskit_serverless}. They propose collecting telemetry data on an ongoing basis to use a machine learning model for predicting future usage and optimizing scheduling.

Reference~\cite{literature_res_23} suggests considering the scheduling time granularity $T_\text{sched}$ in relation to the time a job takes on the quantum computer $T_q$ for the loose integration model. For batch scheduling, $T_\text{sched}$ is in the range of one minute. Jobs on superconducting hardware execute much faster ($T_q \ll T_\text{sched}$). The author argues that, in this case, a low-level queue is required to handle the granularity of quantum jobs independently of the batch scheduler. However, if $T_q \gg T_\text{sched}$, as is the case for neutral atoms, there is a risk of the quantum part blocking the classical resources for too long. In this case, the author suggests using a workflow manager (see Section~\ref{sec:workflow_models}), which does not block classical resources while the quantum part of the application is running.

An additional difficulty in scheduling arises when several quantum computers are available. Even if they are the same model, parameters such as fidelity vary during runtime. This should be considered when assigning the best-fitting resource to a job~\cite{literature_res_309}. It is also possible to divide circuits into smaller subcircuits and run them simultaneously on different quantum computers~\cite{circuit_cutting}. This technique, called circuit cutting, allows large circuits to be run on smaller devices, albeit at the cost of exponential sampling overhead. Although circuit cutting is primarily a compiler task, it must also be considered during scheduling.

The Munich Quantum Software Stack provides support for these scenarios~\cite{literature_res_476}. First, circuits are optimized with a target-agnostic optimizer. Then, subcircuits are generated and assigned to specific hardware. A target-specific optimization stage then further optimizes the circuit based on the chosen hardware, considering the gate set and topology of the device. In the initial version, scheduling is simple, with each subcircuit assigned to the next available device. Reference~\cite{literature_res_345} improves scheduling by considering circuit characteristics, runtime parameters such as queue length and estimated noise, and user preferences. The tool uses these parameters to find an optimal schedule by performing either a meta-heuristic or a reinforcement learning approach. The meta-heuristic approach optimizes for makespan; the reinforcement learning approach also optimizes for low noise. The tool is available on Github~\cite{milq_github}.

The Qonductor~\cite{literature_res_309} scheduler (see Section~\ref{sec:workflow_models} for the corresponding programming model) performs scheduling by formulating an optimization problem that minimizes mean job completion times while maximizing mean fidelity. A genetic algorithm is then used to obtain a Pareto front of solutions. From this front, a single solution can be selected based on the relative importance of job completion times and mean fidelity. The scheduler is regularly invoked to create a schedule for the existing queue. In the example shown in the paper, this occurs either when the job queue reaches 100 elements or after 120 seconds. This scheduling approach relies on the ability to accurately and efficiently predict job completion times and fidelity. Runtime is predicted based on circuit width and depth and the number of two-qubit gate operations using a regression model of previous runs. Fidelity is estimated based on the gates used and the hardware's calibration data.

We conclude that the missing middleware, which was noted in 2016~\cite{literature_res_211}, has since been developed by several different projects. The existing work mainly focuses on scheduling and managing resources. This is because quantum computers are currently scarce and must be utilized as effectively as possible. We found little information about communication protocols in the papers. This might be because each quantum computer is controlled by a classical computer that can communicate with other classical computers via already established classical protocols. So in a way, the title of Reference~\cite{literature_res_23} is right: ``It's all about scheduling''.

\section{Benchmarking}
\label{sec:benchmarking}

Benchmarks are commonly used in computer science and engineering to evaluate the performance of systems. They focus on a component of a system or the entire system and provide a metric that is suitable for comparison with other systems. The metric reflects some aspect of the system, such as computational power or energy efficiency. Benchmarks are critical in the development of new hardware or software to compare different solutions.

Many benchmarks have been developed in the field of HPC, the most famous being the LINPACK benchmark~\cite{linpack}, which is used to rank systems on the TOP500 list~\cite{top500}. LINPACK is often criticized for only representing a system's ability to solve dense linear equations, which is not necessarily representative of real-world applications. It is a good illustration of the advantages and disadvantages of benchmarks: systems can be easily compared, in the case of LINPACK on the basis of a single number, but they only represent a certain aspect of the system. Specialized benchmarks have been developed for quantum computers. Two commonly used metrics are quantum volume~\cite{quantum_volume} and circuit layer operations per second (CLOPS).

For hybrid systems, it is possible to benchmark the classical and quantum parts of the system separately using state-of-the-art methods in the respective field. However, this does not cover the performance of the whole system, since the interaction of both parts is not taken into account. Therefore, it is necessary to develop benchmarks specifically for hybrid systems. Reference~\cite{literature_res_242} gives a general overview of benchmarking for quantum computers, one part of which deals with benchmarking quantum computers at the HPC level, taking into account their integration into HPC systems. The authors note the lack of benchmarks that consider the characteristics of hybrid systems, such as the latency between quantum and classical processors or the efficiency of compilation instructions for the quantum computer. They argue that these aspects are relevant to the performance of applications that users run on hybrid systems.

To cover these aspects, it is possible to perform benchmarks at the application level. In the HPC area, there is the concept of mini-apps, which represent a simplified form of typical applications. In Reference~\cite{literature_res_25}, three mini-apps for hybrid systems are developed: random circuit execution, a variational quantum algorithm, and a quantum machine learning pipeline. These applications are intended to represent typical execution motifs on hybrid systems. The framework is available on Github~\cite{miniapps_github}.

A lower-level benchmark is introduced in Reference~\cite{literature_res_473}, specifically to evaluate the performance of variational algorithms. The authors propose a linear model to estimate the latency to execute a circuit with $n$ shots on the quantum computer and return the results: 
\begin{equation}
    T(n) = T_V + n T_Q,
\end{equation}
where $T_V$ is the step latency and $T_Q$ is the shot latency. The step latency indicates how long it takes to execute a circuit with one shot on the quantum computer. The shot latency describes the increase in latency due to an additional shot. The circuit run for this test is a variation of the log quantum volume. It uses random phase gadgets ($R_Z$ gates with previous and subsequent CNOTs), which represent the typical structure of variational algorithms.

\section{Discussion and Future Topics}
\label{sec:future_topics}

One goal of this paper is to identify gaps and starting points for further research. Various research groups have already worked on many aspects of the topic. From a hardware perspective, we think that on-premises integration is currently the most promising option. Several HPC providers have demonstrated the feasibility of operating quantum computers in close proximity to HPC systems and connecting them via a high-speed interconnect. In addition to lower latency, this setup has the major advantage of potentially improved scheduling. Further reducing latency would require tight integration, which is difficult to achieve and arguable unnecessary since quantum circuit runtimes are much longer than the latency with on-premises integration. However, error correction and classical feedback loops require low latency. We believe these processes will be handled by a special processor close to the hardware rather than the HPC system.

Today's systems integrate with the loose integration model, either remotely or on-premises. We believe that workflow models (see Section~\ref{sec:workflow_models}) are a good starting point for both integration models. Workflow models provide simple access to the quantum computer without requiring the deep adoption of existing HPC tools. Deeper integration can be done step by step subsequently. Most of the tools presented here are open source and easy to experiment with.

Older publications note the lack of middleware and scheduling concepts, whereas newer publications focus more on these topics. We believe the field is currently in a state where many different solutions are being implemented and tested. The next big step should be to standardize protocols and tools to avoid having many singular solutions. In the future, it should be possible to integrate hardware from different vendors into an existing infrastructure without the need for additional software solutions. However, since quantum hardware is rapidly evolving and fault-tolerant hardware has not yet been achieved, it may be premature to standardize.

Many programming models are available, but they focus on circuit-level programming, which is inaccessible to HPC users. Providing higher-level frameworks is difficult, but we believe it will be crucial for the widespread adoption of quantum computing. We also see a lack of universally accepted benchmarks for hybrid systems. Should quantum computers become standard elements of HPC systems in the future, there should be a way to rank the most powerful hybrid systems, similar to the TOP500 list.

\section{Conclusion}
\label{sec:conclusion}

We conducted a structured literature review on the integration of quantum computers into HPC systems. Our analysis is based on 107 publications from 2016 to March 2025. The paper organizes the available literature into seven subtopics. For each topic, we explained the relevant concepts and provided an overview of the available literature, as well as overview tables. This allows for a quick overview as well as a deeper study of individual topics based on the provided literature.

We conclude that substantial work has been done in the field of integrating quantum computers into HPC systems. Nevertheless, we do not expect interest in this topic to decline in the near future, as quantum hardware is becoming increasingly accessible. More and more HPC centers will integrate quantum computers into their systems to prepare for the future of HPC.

\printbibliography

\end{document}